\documentclass[conference]{IEEEtran}
\IEEEoverridecommandlockouts

\usepackage{subfig}

\usepackage{tikz}
\usetikzlibrary{chains}
\usetikzlibrary{fit}
\usetikzlibrary{arrows}
\makeatletter
\pgfdeclareshape{datastore}{
  \inheritsavedanchors[from=rectangle]
  \inheritanchorborder[from=rectangle]
  \inheritanchor[from=rectangle]{center}
  \inheritanchor[from=rectangle]{base}
  \inheritanchor[from=rectangle]{north}
  \inheritanchor[from=rectangle]{north east}
  \inheritanchor[from=rectangle]{east}
  \inheritanchor[from=rectangle]{south east}
  \inheritanchor[from=rectangle]{south}
  \inheritanchor[from=rectangle]{south west}
  \inheritanchor[from=rectangle]{west}
  \inheritanchor[from=rectangle]{north west}
  \backgroundpath{
    \southwest \pgf@xa=\pgf@x \pgf@ya=\pgf@y
    \northeast \pgf@xb=\pgf@x \pgf@yb=\pgf@y
    \pgfpathmoveto{\pgfpoint{\pgf@xa}{\pgf@ya}}
    \pgfpathlineto{\pgfpoint{\pgf@xb}{\pgf@ya}}
    \pgfpathmoveto{\pgfpoint{\pgf@xa}{\pgf@yb}}
    \pgfpathlineto{\pgfpoint{\pgf@xb}{\pgf@yb}}
 }
}
\makeatother

\title{DePIN: A Framework for Token-Incentivized Participatory Sensing}

\author{Michael T. C. Chiu\thanks{M. T. C. Chiu (michael@wihi.cc) is with WiHi.} \and Sachit Mahajan\thanks{S. Mahajan (sachit.mahajan@gess.ethz.ch) is a Lecturer of Computational Social Science at ETH Zurich.} \and Mark C. Ballandies\thanks{M. C. Ballandies (mark@wihi.cc) is with WiHi.}  \and Uro\v{s} V. Kalabi\'{c}\thanks{U. V. Kalabi\'{c} (uros@wihi.cc) is with WiHi.}}

\begin{document}

\maketitle

\begin{abstract}
There is always demand for integrating data into microeconomic decision making. Participatory sensing deals with how real-world data may be extracted with stakeholder participation and resolves a problem of Big Data, which is concerned with monetizing data extracted from individuals without their participation. We present how Decentralized Physical Infrastructure Networks (DePINs) extend participatory sensing. We discuss the threat models of these networks and how DePIN cryptoeconomics can advance participatory sensing.

\end{abstract}

\section{Introduction}

The world is interconnected and advancements in information and communications technology are readily improving information transfer between interconnections. The world economy is complex and 
improvements in microeconomic data-sharing are, as a matter of course, leveraged to remove market inefficiency and, for example, improve price discovery or even improve liquidity through more efficient use of leverage. 
Economic inefficiencies are prone to exploitation for financial gain, so
there is always demand for real-world data that can be used to advise microeconomic decisions.

Distributed ledger technology (DLT), and the closely associated concepts of blockchain and cryptocurrency, allows for the democratization of information exchange by enabling the establishment of a source of truth, \textit{i.e.}, the ledger, which, when adequately regulated through the use of incentivization schemes, can closely align the state of the ledger with the state of affairs reflected by whatever part of reality a ledger is designed to reflect. In this way, a ledger may reflect either physical data, financial data, or both; from a technical perspective, the specific type of data is irrelevant. However, from the perspective of design, what the data represent is important, 
\textit{i.e.}, technical aspects of DLT are of less concern to data analysts as opposed to what the ledgers themselves represent and how their contents represent real-world value.

The concept of web3 is one of a decentralized World Wide Web where the transfer of value is governed through the use of DLT and cryptocurrenices. The preceding concept of web2 is one where data are typically siloed and the monopsonic pricing power held by larger entities is wielded against individuals to extract value from their data while offering significantly less in return. A conceptual outgrowth of web2 has been the advent of Big Data, but big data is not necessarily \emph{good} data; when  data  collected surpasses processing power available, it results in sampling biases \cite{helbing2015thinking}. This is where web3 shows greater promise.  

In particular, in web3 it is important to consider incentive alignment and offer more equitable terms to individuals. The use of cryptocurrencies allows for this because they are permissionless, borderless and, depending on the DLT used, cheap. However, using DLT solely for the purpose of value transfer via cryptocurrency is not on its own an insufficient use of DLT since DLT can be so much more\cite{dapp2021finance}: No longer is it necessary to separate finance from data at the level of transaction processing. The web3 methodology allows the possibility of merging data into financial processes in more complex ways that unlock the ability to extract good data at fair market value.
Compare Figures 1a and 1b. The first figure represents a web2 design methodology, where money, in this case cryptocurrency, is exchanged for some data between two entities. The second figure represents a web3 design methodology, where both money and data are interlinked in a complex way and it is more difficult to identify value flows because an individual entity may be both customer and provider simultaneously.

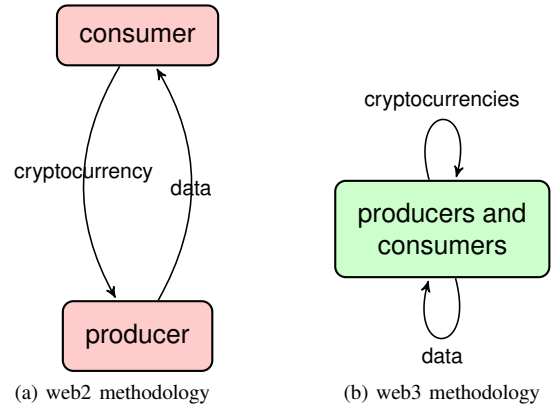
\begin{figure}[tp]
\centering
\subfloat[web2 methodology]{
\begin{tikzpicture}[
  font=\sffamily,
  rect/.style={draw,thick,rounded corners,fill=red!20,inner sep=.3cm},
  to/.style={->,>=stealth',shorten >=1pt,semithick,font=\sffamily\footnotesize},
  every node/.style={align=center}]

\node[rect] (participant) at (0,-2) {producer};
\node[rect] (consumer) at (0,2) {consumer};

\draw[to] (consumer) edge[bend right=30] node {cryptocurrency \\} (participant);
\draw[to] (participant) edge[bend right=30] node {\\ data} (consumer);
\end{tikzpicture}
}
\hspace{0.45in}
\subfloat[web3 methodology]{
\begin{tikzpicture}[
  font=\sffamily,
  rect/.style={draw,thick,rounded corners,fill=green!20,inner sep=.3cm},
  to/.style={->,>=stealth',shorten >=1pt,semithick,font=\sffamily\footnotesize},
  every node/.style={align=center}]

\node[rect] (self) {producers and \\ consumers};

\draw[to] (self) edge [loop above] node {cryptocurrencies} (self);
\draw[to] (self) edge [loop below] node {data} (self);
\end{tikzpicture}
}
\caption{Conceptual differences between web2 and web3}
\label{fig:web2-web3-diffs}
\end{figure}

This is where the concept of Decentralized Physical Infrastructure Networks (DePINs) becomes important. DePINs are a novel way of organizing physical infrastructure \cite{ballandies2023taxonomy}, such as the energy grid, that leverages DLT to unlock novel ways of sourcing data, consuming data and services and building the overall platform. To expand on the framework of Figure 1b, we present in Figure 2 a schematic of how DePINs commonly work: a platform, ideally running on-chain via smart contracts, serves as both an ingress and egress point for both cryptocurrencies and data; the data flows from producers to consumers, whereas cryptocurrencies  flow between producers (who are paid in cryptocurrency and exchange it for their preferred currency), consumers (who pay in cryptocurrency and procure it with their preferred currency), maintainers (who, like producers, are paid in cryptocurrency), and speculators (who perform price discovery). The point here is that individuals may, at any point in time, be producer, consumer, maintainer, speculator, or any combination of the foregoing.

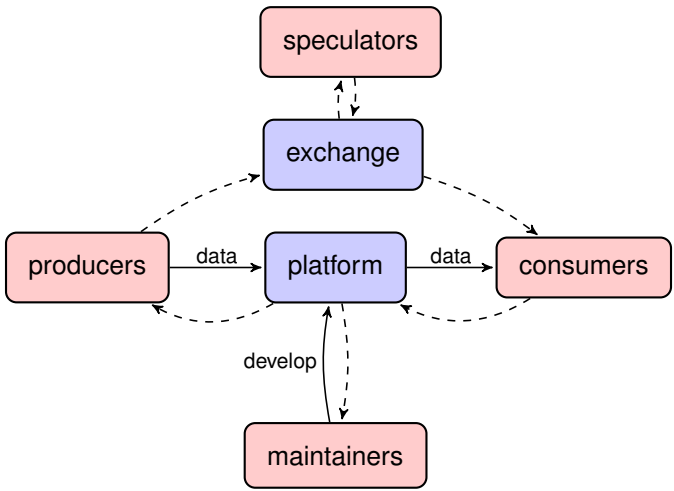
\begin{figure}[t]
\centering
\begin{tikzpicture}[
  font=\sffamily,
  rect/.style={draw,thick,rounded corners,fill=blue!20,inner sep=.3cm},
  participant/.style={draw,thick,rounded corners,fill=red!20,inner sep=.3cm},
  to/.style={->,>=stealth',shorten >=1pt,semithick,font=\sffamily\footnotesize},
  too/.style={->,>=stealth',shorten >=1pt,semithick,dashed,font=\sffamily\footnotesize},
  every node/.style={align=center}]

\node[rect] (platform) at (-1.7,0) {platform};
\node[participant] (providers) at (-5,0) {producers};
\node[participant] (consumers) at (1.6,0) {consumers};
\node[rect] (exchange) at (-1.6,1.5) {exchange};
\node[participant] (speculators) at (-1.5,3) {speculators};
\node[participant] (maintainers) at (-1.7,-2.5) {maintainers};

\draw[too] (platform) edge[bend left=30] node {} (providers);
\draw[too] (consumers) edge[bend left=30] node {} (platform);
\draw[to] (providers) edge node {data\\} (platform);
\draw[to] (platform) edge node {data\\} (consumers);
\draw[too] (providers) edge[bend left=10] node {} (exchange);
\draw[too] (exchange) edge[bend left=10] node {} (consumers);
\draw[too] (speculators) edge[bend left=10] node {} (exchange);
\draw[too] (exchange) edge[bend left=10] node {} (speculators);
\draw[too] (platform) edge[bend left=10] node {} (maintainers);
\draw[to] (maintainers) edge[bend left=10] node {develop\hspace{0.45in}} (platform);
\end{tikzpicture}
\caption{Schematic of value transfer within a DePIN ecosystem (dashed lines represent cryptocurrency flows)}
\label{fig:depin-scheme}
\end{figure}

More generally, DePINs can be considered a part of the Internet of Things (IoT) \cite{ballandies2023taxonomy}, in which extending internet connectivity to a wide range of physical devices and everyday objects, enabling them to collect and exchange data and enhance automation, efficiency, and data-driven decision-making in various domains. 
An older concept than that of DePIN, related to IoT, is that of participatory sensing \cite{burke2006participatory}. Participatory sensing recognizes that inexpensive sensing devices, such as smartphones with their ubiquity and persistent Internet connectivity, may be leveraged to provide valuable real-world insights. For example, participatory sensing has been used to monitor air pollution \cite{mendez2011}, 
noise pollution \cite{schweizer2011noisemap}, and
street illumination levels \cite{middya2021citylightsense};
crowd-source bargin-hunting goods \cite{deng2009livecompare},
the detection of pot-holes \cite{patra2021potspot}, and
determination of thermal comfort \cite{erickson2012thermovote}. However, 
the data provided by inexpensive sensors is typically of limited value when compared to professional-grade sensing technology. In this sense, participatory sensing does not offer much of a solution to the inadequate-data problem of Big Data, described previously.

Despite significant advancements in participatory sensing, a multitude of challenges remain, particularly in ensuring high-quality data. These challenges extend beyond technical issues like device calibration and sensor drift \cite{can2016cross}, which impact accuracy and reliability. They also encompass aspects such as participant motivation, data privacy concerns, and the management of heterogeneous data sources \cite{mahajan2023democratizing}. Additionally, the variability in participant engagement and the potential for biased data collection due to uneven geographic or demographic representation pose significant hurdles. In addressing these challenges, the DePIN concept and the broader field of cryptoeconomics are reshaping participatory sensing. They offer frameworks for scaling up these initiatives, ensuring data integrity, and providing incentives for participation. This shift towards a more structured and incentivized model enables participatory sensing to be effectively implemented on a larger scale, improving its potential for application and impact.

In this paper, we present how DePIN as a framework, and cryptoeconomics more generally, reformulate participatory sensing so that it becomes applicable at scale. We present DePIN-related approaches to hardware design, software architecture, and incentivization mechanism that may advance the field of participatory sensing to reach its full potential.

We begin by dicussing the lack of related work; in the following section, we describe the technical problems faced by participatory sensing networks; in the section following that, we describe how DePINs may be used to provide a solution.

\subsection{Related Work}
DePIN is a novel concept undergoing rapid development. As such, the authors are confident in remarking that there is no work directly related to studying the relationship between participatory sensing and DePIN. Work that is indirectly related includes the study of the advantages of token-incentivized systems over traditional approaches 
\cite{malinova2023tokenomics} and exploring improvements in DePIN architectures that can potentially enable the next level of scalability of DePIN systems \cite{fan2023towards}.

\section{Challenges in Participatory Sensing}

Participatory sensing often adopts an open and permissionless approach, enabling widespread and inclusive participation. While this is a key strength of the participatory model, openness also presents unique challenges, especially with regards to incentivizing participation. As participation is open to all, any system of rewards intended to encourage genuine contributions can simultaneously attract malicious actors,
whose readiness to provide false data scales in proportion to the size of reward.

Until the advent of Bitcoin, a similar problem existed in money transmission over the Internet, so the Proof-of-Work (PoW) consensus mechanism \cite{nakamoto2008bitcoin} was introduced to enable trustless consensus. 
However, although PoW provides an effective
defense against Sybil attacks, 
it is impractical to run it at the level of a sensor. This is because, regardless of the cost and quality of the sensing technology, PoW will prove uneconomical.
Furthermore, even in the absence of monetary rewards, participatory sensing networks may experience Sybil attacks\cite{verchok2020hunting} and it is therefore necessary that such networks implement defenses that prevent these. In the following, we describe the types of defenses that may be implemented.

\subsection{Hardware-Based Defenses}

%
%


The most common approach to prevent Sybil attacks in participatory sensing is hardware based, 
typically requiring the use of a Trusted Platform Module (TPM)
to guarantee trust in a sensor\cite{saroiu2010sensor}. A TPM is a
tamper resistant chip separate from a sensing device's processor with 
 the capability to access protected memory and registers,
generate random numbers, seal data to system state, and manage
and store cryptographic keys securely\cite{ezirim2012trusted}. 
%
%
%
TPMs enable trust in a system in a number of ways.
For example, TPMs enable measured boot, a boot protocol
in which every layer of the firmware is ``measured'', typically by hashing the firmware, before 
being loaded and securely stored for later verification. 
TPMs also enable Remote Attestation (RA), a challenge-response protocol 
between an untrusted prover, \textit{i.e.}, a sensor, attempting
to prove that it has 
determined the state correctly, and a verifier, \textit{i.e.},
the network, attempting to determine the 
trustworthiness of the untrusted prover \cite{banks2021remote}. 
Nevertheless, incorrect usage of TPM APIs can render TPMs useless and this is not an uncommon occurrence\cite{wan2020}.
Furthermore, although TMPs
are becoming increasingly
ubiquitous, being shipped with commonly used TPM-enabled microcontrollers like the Raspberry Pi\cite{pinto2019virtualization}, requiring
users to install 
custom software that can access the TPM can prove difficult for the purposes of both on-boarding participants, especially in the early stages of network growth where any data source is welcome, and on-boarding hardware manufacturers, who could be resistant to allow their devices to be tampered with.

\subsection{Server-Based Quality Assurance}

Data received from a sensor running the correct firmware is not neccessarily either of quality or trustworthy. 
Apart from maliciousness, this
may be due to improper sensor setup or some fault in either hardware, firmware, or even the communication channel. 
Data verification in participatory sensing systems refers to the problem of 
ensuring data accuracy, removing outliers, data completeness, and consistency, data integrity, and spatial and temporal validation \cite{chen2018pm2.5}. Approaches include: spatial interpolation\cite{middya2021spatial}, inverse distance weighting\cite{bilonick1991introduction}, Kriging\cite{aumond2018kriging}, deep neural nets (and associated preprocessing algorithm)\cite{chang2023deep}, cross validation\cite{lou2019iotdatacv}, unsupervised learning\cite{banerjee2018unsupervisediot}, and the use of optimization\cite{restuccia2018first}. These approaches are based in software run on the server receiving the data and, in general, these approaches compare sources of data between to each other to determine whether data from particular sensors surpasses some quality threshold. 

\subsection{Mechanism Design}

The main weakness of participatory sensing is that of adequately incentivizing participants.
Conventional game-theoretic analysis in the context of participatory sensing, such as framing the problem as a Stackleberg game 
or reverse auction where the  user who bids with the least reward obtain rights to participate in a sensing task \cite{yang2012}, has shown 
that financial incentives are effective at incentivizing participants to perform tasks in proportion to the quantity of data shared \cite{christin2013s}. Somewhat remarkably, 
it has also been shown that fiat-based financial incentives, such as the use of micro-payments, have resulted in participants losing focus a short while after taking up tasks \cite{reddy2010} and that the impact on quality of shared information may be negative \cite{ballandies2022to} since 
intrinsic motivation can be crowded out through the use of fiat-based incentives\cite{osterloh2000motivation}. 
A focus on long-term incentivization, however, increases social welfare \cite{gao2015longterm} but a further downside to pure finanical incentive mechanisms, besides the crowding out of intrinsic motivation, is that it requires the use of actual money, which can result be costly for the network operator. 

For these reasons, alternatives to financial incentivization have been explored, including gamification \cite{ueyama2014gamification}, reputation \cite{yu2019participant,huang2010you}, and intrinsic motivation \cite{christin2013s} mechanisms. 

\section{DePIN: A Framework for Token-Incentivized Participatory Sensing}
\label{sec:depin-part-sense}

Decentralized Physical Infrastructure Networks (DePINs) use cryptoeconomics such as token-incentives, decentralized governance, and distributed ledger technology to solve many of the same challenges faced by participatory sensing networks and, as such, enable their scaling. While still in its infancy, DePINs can be defined as decentralized networks that utilize cryptoeconomics to incentivize participants to build physical infrastructure or procure resources that stem from a physical asset. Two widely accepted examples of DePIN networks are: Helium and Filecoin.

DePINs have all the elements of a participatory sensing network: participants, on a large-scale, contribute to the functioning of the network by providing resources. An improvement over participatory sensing networks, however, is that incentivization, in one form or another, is tied to the network token: the monumental success of DePINs such as Helium and Filecoin are a testament to this. The network token enables tokenomics and other game-theoretic mechanisms to not only incentivize participation but to also disincentivize malicious behavior. For this reason, DePINs should be seen as a framework for token-incentivized participatory sensing.

In the following, we begin by introducing the threat model of DePIN, and then present how cryptoeconomic mechanisms may be used to mitigate these threats. In particular, we make the case that cryptoeconomic mechanisms can be a robust approach to disincentivize participants from carrying out attacks on the network.

\subsection{Threat Model and Sensor Node Security}
\label{sec:depin-threat-model-node-sec}

Determining trustworthiness of contributed data is important for both participatory sensing networks as well as DePINs. 
In the case of the latter, the fact that nodes are incentivized for long-term participation implies that they have a higher incentive to act maliciously. Morevover,
discouraging individual nodes from providing malicious data is challenging because it is non-trivial to determine the relationship between quality of the received data stream and potential reward. 
The need to prevent malicious behavior is not restricted to open-hardware use cases since,
although restricting the specific hardware that may register on the network may help with preventing Sybil attacks, the fact remains that a malicious participant may tamper with the environment itself to provide a false, \textit{i.e.}, more beneficial to the participant, sensor reading.




In the following, we describe threats on a per-node basis. 
We begin by defining a DePIN sensor node and, by extension, participatory sensing node.

\paragraph*{Sensor Node.}
A sensor node within a DePIN (resp. participatory sensing network) is a physical hardware device 
having the following components:
\begin{enumerate}
\item a processing unit (CPU)
\item writeable memory storage unit (RAM)
\item non-writeable memory (ROM)
\item sensing (measurements) peripherals 
\end{enumerate}
This definition of sensors allows for a broad range of
hardware, ranging from microcontroller-like devices with 
low computational capabilities to more powerful Raspberry Pis,
wherein, in both cases, the peripherals provide sensing or measurement capabilities. There are three main type of threats that sensor nodes face.

\subsubsection{Device Threats}
Device-level threats include both hardware and software threats. 
Hardware threats are more commonly known as firmware-level threats
since firmware is the crucial low-level piece of software that is responsible
for booting a device or communicating with peripherals (drivers).
It is almost impossible to completely defend against firmware threats during runtime \cite{shah2020survey}
but this concern can be greatly alleviated by strengthening a chain-of-trust \cite{zimmer2016establishing}. 
Software threats are the more traditional and well-known
type of attacks that occur during runtime \cite{or2019dynamic}. In general, software-level
threats are addressed by some form of ``Remote Attestation''. However, 
attestation for less powerful devices such as sensors   are not ideal as they require more power \cite{ammar2020simple} or that the sensor be briefly paused \cite{dushku2020sara}.
We note that, as with all electronics, physical access to hardware
can bypass all firmware and software protections \cite{fu2018}. Hardware approaches to ensuring sensor node security 
either reduce the potential size of the network, by requiring additional functionality such as, for example, 
trusted hardware, or are not sufficient on their own to address the 
threats that a DePIN sensor network face.

\subsubsection{Network Threats}

Sensors communicate with the broader network over the 
Internet 
via APIs. A malicious actor can bypass an actual 
physical device by emulation 
and falsify measurements directly
to the network via the API boundary \cite{wang2018ghost}. 
Note that closed-hardware solutions do not adequately address this class of threats since
they greatly reduce the potential size of the network. 

\subsubsection{Sensor Environment Threats}

An important class of threats unique to decentralized sensor networks, 
whether classical participatory sensing networks or DePIN sensor 
networks, are attacks where the malicious participant alters the
physical environment of the sensor node or introduces an artificial
element in the sensor node's physical environment. For example, 
a malicious actor can place a sensor at a sub-optimal location for 
measurement in order to record data that might be viewed as 
and thus 
valuable by the network. Similarly, a malicious actor
can artificially create or modify the sensor environment to achieve the same effect. 
Sensor environment threats are, arguably, the main threats to
DePINs without an adequate way to address 
this type of threat.

%
%

\subsection{Cryptoeconomic Mechanisms}
\label{sec:depin-sensor-incent-penal}
Cryptoeconomics is utilized by DePINs to mitigate the threats described above and 
challenges of participatory sensing. Cryptoeconomics consists of, amongst other things, tokenomics, governance and DLT \cite{voshmgir2019foundations}.


\subsubsection{Tokenomics}

\paragraph{Tokens}

Monetary incentives are effective in promoting active and substantial participation in both participatory sensing \cite{christin2013s} and DePINs \cite{jagtap2021federated}. The key distinction lies in the reward types: DePINs use network-specific tokens as incentives, whereas traditional participatory sensor networks, if they offer incentives at all, typically provide cash or non-exchangeable rewards like point systems.
In contrast to monetary incentives, token-based incentives allow for  designs that increase the intrinsic motivation of network participants to contribute in quality as well as quantity, 
a major limitation of fiat-only approaches being that they often crowd out intrinsic motivation \cite{ballandies2022incentivize}. For instance, token incentives can represent ownership or reputation, potential drivers of intrinsic motivation \cite{kuwabara2015reputation}.
For this, a token can be constructed from a large design space \cite{ballandies2023taxonomy}: System designers can define how a token may i) be burned, removing units from circulation; ii) be transferred; iii) be capped in supply, iv) be premined, v) be limited in fungibility, vi) have a source of value; and vii) have a creation mechanism bound to concrete actions, \textit{e.g.}, the sharing of sensor data.

Moreover, token-based rewards shift future earnings to the present, offering immediate financing for DePIN systems. This incentivizes participation, overcomes budget constraints and enables the creation of networks that might not be feasible without such incentives \cite{jagtap2021federated}. Thus, DePINs can bootstrap more effectively than traditional, non-incentivized network development.

\paragraph{Multi-Token Models}

The flexibility of token-based approaches allows a system designer to deploy more than one token and in this way span a multi-dimensional incentivization space that can result in an improved calibration and thus resilience of a cryptoeconomic system \cite{dapp2021finance}.
The most prevalent multi-dimensional incentivization approach in DePINs is the burn-and-mint model, which effectively aligns the token's value with the network's service value \cite{kalabic2023burn}. This dual-token system consists of a 'value' token, created from nothing to reward nodes for their services, and a 'utility' token, with a fixed fiat value for buying services. The value token is traded openly, while the utility token is acquired by destroying an equivalent amount of the value token. 
Often the dimension of these token models are increased over a systems lifetime, as it is for instance observed in Helium. 

\paragraph{Game Theory}

On top of these token models, several game-theoretical mechanism can been applied to provide further incentives to contribute in terms of quantity and more importantly in quality to a DePIN system,
such as staking \cite{kraner2023agent}, vesting \cite{schar2021decentralized}, or bonding curves \cite{zargham2020curved}.

\paragraph{Participatory Governance}
DePINs can scale to large interdependent networks of a diverse set of stakeholders. 
These techno-socio-economic networks are complex systems \cite{voshmgir2019foundations} where traditional governance and control mechanisms often fail, such as in the case of sustainability and resilience \cite{helbing2021networked}.
Hence, bottom-up, decentralized mechanisms are increasingly used to navigate these complex systems and have been shown to control and calibrate them more effectively \cite{helbing2023democracy}.
An expression of this trend is the emergence of decentralized autonomous organizations (DAOs) that combine collective intelligence, digital democracy and self-organization \cite{helbing2023democracy} to navigate complex blockchain systems. 
DAOs comprise two main elements: the community and the organization \cite{ospina2023daocommunity}. The community consists of individuals united by relationships and a common identity, each with their own goals like investment returns or enjoyable experiences. A DAO forms when these members collaborate to fulfill a shared vision that aligns with their personal aims. This structure offers a sense of belonging and purpose, addressing this key shortcoming of earlier participatory sensing campaigns \cite{balestrini2015participatory}.
The governance of DePIN networks is often centralized \cite{ballandies2023taxonomy}, which can result in rent extraction or hold-up problems \cite{goldberg2023metaverse} and generally undermine the decentralization of a DePIN. 

\subsubsection{Distributed Ledger Technology} 
Several concepts from DLT can be potentially be utilized in DePIN to mitigate the illustrated challenges. For example, a useful work type of consensus algorithm, as utilized for instance in Filecoin, can prevent Sybil attacks and can guarantee quality of service, e.g. trustworthy sensor data. Nevertheless, no generalizable solution to date has been found for DePIN networks.
Furthermore, security and privacy are major concerns in participatory sensing \cite{karim2020big} which can be mitigated by using DLT \cite{cheng2020reputation}, e.g. the immutable storage or zero-knowledge proofs.
Finally, decentralized identities could facilitate better control of DePIN contributors about their data, another limitation of participatory sensing \cite{karim2020big}.











\section{Conclusion}
\label{sec:concl}
There is always demand for integrating data into financial decision making. Large data sets are valuable but when the origins of data are spread across many stakeholders, the data becomes difficult  to extract. The field of participatory sensing is concerned with finding ways to share and extract such data.  

In this paper, we showed that cryptoeconomics applied, through the framework of DePIN, holds great promise for tackling the challenges of low participation and insufficient data quality in participatory sensing networks.
We demonstrated the next stage in participatory sensing, proposing directions on how the field may be improved through the integration of cryptoeconomic mechanisms through the use of 
Decentralized Physical Infrastructure Networks (DePINs). 
We presented threats faced by participatory sensing networks as well as DePINs, and how these threats may be addressed.



\bibliographystyle{ledgerbib}
\bibliography{nakamoto}

\end{document}